\documentclass[aps,prl,reprint,groupedaddress,intlimits]{revtex4-1}

\usepackage{amssymb}
\usepackage{amsthm}
\usepackage{graphicx}
\usepackage{dcolumn}
\usepackage{bm}
\usepackage{amsfonts}
\usepackage{amsmath}
\usepackage{amssymb}
\usepackage{color}
\usepackage{natbib}
\usepackage{float}
\usepackage{upgreek}

\begin{document}


\title{Spontaneous polarization and locomotion of an active particle with surface-mobile enzymes}


\author{Marco De Corato}
\email{mdecorato@ibecbarcelona.eu}
\affiliation{Institute for Bioengineering of Catalonia (IBEC), The Barcelona Institute of Science and Technology (BIST), Baldiri Reixac 10-12, 08028 Barcelona Spain\looseness=-1}

\author{Ignacio Pagonabarraga}
\affiliation{Departament de F\'{i}sica de la Mat\`{e}ria Condensada, Universitat de Barcelona, C. Mart\'{i} Franqu\`{e}s 1, 08028 Barcelona, Spain \\
University of Barcelona Institute of Complex Systems (UBICS), Universitat de Barcelona, 08028 Barcelona, Spain \\
CECAM, Centre Européen de Calcul Atomique et Moléculaire, École Polytechnique Fédérale de Lasuanne (EPFL), Batochime, Avenue Forel 2,1015 Lausanne, Switzerland\looseness=-1}

\author{Loai K. E. A. Abdelmohsen}
\affiliation{Department of Biomedical Engineering, Department of Chemical Engineering and Chemistry, Institute for Complex Molecular Systems Eindhoven University of Technology, P.O. Box 513, Eindhoven MB 5600, The Netherlands \looseness=-1}

\author{Samuel S\'{a}nchez}
\affiliation{Institute for Bioengineering of Catalonia (IBEC), The Barcelona Institute of Science and Technology (BIST), Baldiri Reixac 10-12, 08028 Barcelona Spain \\
Instituci\'{o} Catalana de Recerca i Estudis Avan\c{c}ats (ICREA), Pg. Llu\'{i}s Companys 23, 08010 Barcelona, Spain\looseness=-1}

\author{Marino Arroyo}
\email{marino.arroyo@upc.edu}
\affiliation{Universitat Polit\`{e}cnica de Catalunya--BarcelonaTech, 08034 Barcelona, Spain\\
Institute for Bioengineering of Catalonia (IBEC), The Barcelona Institute of Science and Technology (BIST), Baldiri Reixac 10-12, 08028 Barcelona Spain \\
Centre Internacional de M\`etodes Num\`erics en Enginyeria (CIMNE), 08034 Barcelona, Spain
\looseness=-1}







\begin{abstract}
We examine a mechanism of locomotion of active particles whose surface is uniformly coated with mobile enzymes. The enzymes catalyze a reaction that drives phoretic flows but their homogeneous distribution forbids locomotion by symmetry. We find that the ability of the enzymes to migrate over the surface combined with self-phoresis can lead to a spontaneous symmetry breaking instability whereby the homogeneous distribution of enzymes polarizes and the particle propels. The instability is driven by the advection of enzymes by the phoretic flows and occurs above a critical P\'{e}clet number. The transition to polarized motile states occurs via a supercritical or subcritical pitchfork bifurcations, the latter of which enables hysteresis and coexistence of uniform and polarized states.  
\end{abstract}

\maketitle



Eukaryotic cells and bacteria use chemical energy to move in various environments. Mimicking living cells and given the availability of chemical energy in the environment, artificial colloidal particles can be designed to self-propel through surface chemical reactions \cite{paxton2004catalytic,howse2007self}. Besides serving as a model system to explore collective nonequilibrium phenomena \cite{bechinger2016active}, several technological applications have been envisaged for these active particles: from biomedical \cite{hortelao2018targeting,tang2020enzyme,hortelao2020monitoring,tang2020enzyme} to environmental remediation \cite{parmar2018micro}. To achieve self-propulsion, different mechanism have been proposed such as diffusiophoresis \cite{sharifi2013diffusiophoretic}, thermophoresis \cite{bregulla2019flow,de2019self}, momentum exchange \cite{eloul2020reactive}, release of ions \cite{de2020self} and liquid-liquid phase separation \cite{buttinoni2012active}.

Regardless of the mechanism, a requirement of self-propulsion is symmetry-breaking. This has been achieved by hard-wiring onto the material particle an asymmetric shape \cite{michelin2015autophoretic,michelin2017geometric,varma2018clustering,baker2019shape} or an asymmetric catalytic reaction rate \cite{golestanian2005propulsion}, both of which pose manufacturing challenges particularly at smaller scales. While built-in asymmetry is intrinsic to flagellates and other  microorganisms, animal cells have the ability to dynamically develop self-polarization of their active cytoskeleton, thereby switching between quiescent and motile states  \cite{RUPRECHT2015673,Bergert:2015aa,PhysRevLett.116.028102}. If polarization is not built-in but is instead an emergent response that can be triggered on demand, this may lead to tunable, adaptable, and more easily produced self-propelled particles. 

\begin{figure}[h!]
\centering
\includegraphics[width=0.5\textwidth]{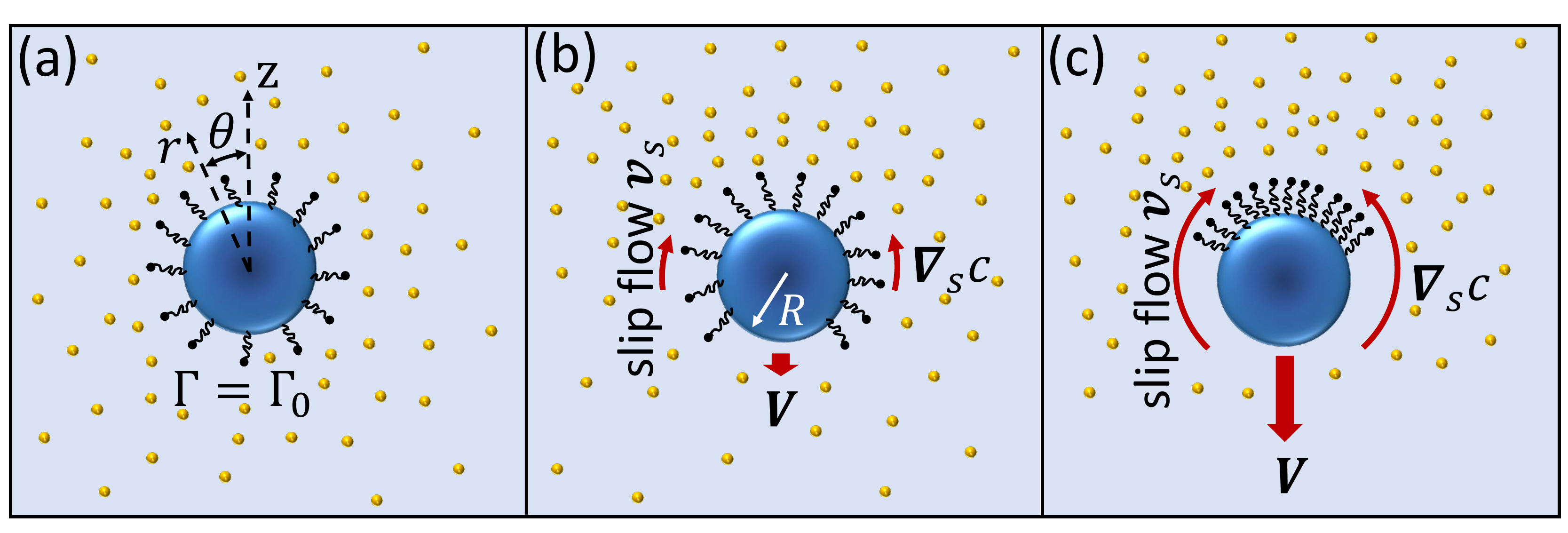}
\caption{Self-propulsion mechanism for a colloidal particle (blue) covered by mobile enzymes (black) in a suspension of molecules catalyzed by the enzymes (yellow). (a) In an unpolarized state, the laterally-mobile enzymes are homogeneously distributed on the surface to maximize entropy and  $\Gamma=\Gamma_0$. (b) A perturbation of the distribution of enzymes leads to an inhomogeneous reaction rate, which generates an imbalance of product on the two sides of the particle. The gradient of product results in phoretic flows that transport more enzymes. Upon a perturbation the system can evolve toward (c), a self-sustained polarization of the enzyme distribution, or go back to (a), a homogeneous state.}
\label{fig1}
\end{figure}

Here we propose a novel self-propulsion strategy based on enzyme catalysis \cite{patino2018fundamental,patino2018influence,ghosh2019motility,somasundar2019positive,arque2019intrinsic} that does not require a built-in asymmetric catalytic reaction. Rather than using a fixed catalyst, we consider chemically-active colloids coated with mobile enzymes. To maximize the entropy, these mobile molecules will tend to homogeneously distribute on the particle's surface. We hypothesize that spontaneous polarization and propulsion may arise as an advective instability driven by the interplay between the surface mobility of enzymes, a chemical reaction and self-diffusiophoresis, as sketched in Figure \ref{fig1}. The mechanism studied here is related to a distinct symmetry-breaking instability in the bulk around isotropic catalytic particles leading to sustained motion, which however requires rather specific experimental conditions \cite{michelin2013spontaneous,schmitt2013swimming,de2013self,izri2014self,maass2016swimming,michelin2020spontaneous}.

To examine this idea, we consider a spherical particle of radius $R$ that is covered by enzymes and is suspended in a liquid, see Fig. \ref{fig1}(a). The enzymes are constrained to the surface of the particle but are free to move laterally. Migration of macromolecules over a surface occurs frequently in nature, a notable example being membrane-associated proteins \cite{tanaka2007frictional,goychuk2019protein,tozzi2019out}.  
We assume that the enzymes are much smaller than the particle, which is the case for micron-sized particles and nanometer-sized enzymes. We thus describe the enzymes through their time-dependent surface number density, $\Gamma$. We neglect thermal fluctuations and the deformation of the surface, which may be important for large and flaccid membranes \cite{gupta2015autonomous}.

In the presence of their substrate, the enzymes catalyze a reaction that releases a product species. We assume that the reaction rate is proportional to the local number density of enzymes $k_{\text{cat}} \Gamma $, with $k_{\text{cat}}$ the turnover rate of the enzyme. This simple choice is pertinent when the substrate concentration is much larger than the Michaelis-Menten constant of the enzyme. The product is released from the surface of the particle and then quickly diffuses in the bulk so that advection is negligible \cite{moran2017phoretic}. It follows that its distribution relaxes instantaneously to steady state after any change of the enzyme distribution. Under these assumptions, the balance of the number density of the product, $c$, satisfies the Laplace equation,
\begin{equation}\label{Laplacedim}
D \, \boldsymbol{\nabla}^2 c = 0 \, \, ,
\end{equation}
with $D$ the diffusion coefficient. At $r \rightarrow \infty $ the concentration of the product is kept constant at $c=0$ and the surface flux of product at $r=R$ is proportional to the reaction rate $-D \, \boldsymbol{\nabla}c\cdot \boldsymbol{n}=k_{\text{cat}} \Gamma $, with $\boldsymbol{n}$ the unit outer normal vector to the surface of the sphere. The main difference with previous models considering chemically active colloids is that the enzymes are mobile and therefore $\Gamma$ can change in space and time.

The reaction product interacts with the surface of the particle through a short-range potential and its gradients along the particle's surface generate lateral gradients of pressure within a thin boundary layer next to the surface \cite{anderson1989colloid,moran2017phoretic}. The size of the boundary layer depends on the details of the product-surface interactions but its thickness is usually in the order of a few nanometers \cite{sharifi2013diffusiophoretic}. The pressure gradient inside the boundary layer is balanced by the viscous shear stress resulting in an apparent slip velocity that develops over a few nanometers from the surface. This mechanism is effectively described through a slip velocity at the surface of the particle that is proportional to the surface gradient of product and given by $\boldsymbol{v}_s = b \, \boldsymbol{\nabla}_s c$, where $b$ is the phoretic mobility coefficient depending on the details of the product-wall interactions \cite{anderson1989colloid}, which we assume to be constant. 
Attractive interactions lead to a negative $b$, while repulsive interactions lead to a positive $b$. 
Finally, for vanishing fluid inertia, the particle velocity is opposite to the surface-average of the slip velocity $\boldsymbol{V} = - \langle \boldsymbol{v}_s \rangle$, where $\langle\;\rangle$ denotes the average over the surface \cite{anderson1989colloid}. 

The enzymes are transported along the surface by diffusion and by the local slip velocity. In a reference frame attached to the center of the particle, their distribution satisfies
\begin{equation}\label{enztransport}
\frac{\partial \, \Gamma}{\partial \, t} = -\boldsymbol{\nabla}_s \cdot \left( \boldsymbol{J}_s+ f \, \boldsymbol{v}_s \Gamma \right) \, \, ,
\end{equation}
where $\boldsymbol{J}_s$ is the diffusive flux of enzymes and $f \, \boldsymbol{v}_s \Gamma$ is flux of enzymes driven by the local slip flow. Since the phoretic velocity goes from zero at the surface to $\boldsymbol{v}_s$ over a few nanometers, we assume that the enzymes are advected by an effective velocity, $f \, \boldsymbol{v}_s $, that is a fraction of the slip velocity $\boldsymbol{v}_s$ observed far from the particle surface. The dimensionless coefficient $f$ takes values between zero and one.   

The diffusive flux of enzymes is proportional to the gradient of their chemical potential, $\mu$, along the surface $\boldsymbol{J}_s = -D_s \Gamma \, \boldsymbol{\nabla}_s \mu /k_BT$, with $D_s$ the surface diffusion coefficient of the enzymes, $k_B$ the Boltzmann constant and $T$ the absolute temperature. The diffusion of enzymes and proteins along membranes is usually much slower than that of small molecules in a liquid $D_s \ll D$ \cite{jacobson1987lateral}, thus neglecting the advective transport of $c$ in the bulk but considering it on the surface is justified.
We assume that the chemical potential that drives the diffusive flux derives from the Flory-Huggins free energy, $\mu = k_BT \log \Gamma /\left(\Gamma_\infty-\Gamma \right) + \chi \, \Gamma -\Lambda \boldsymbol{\nabla}_s^2 \Gamma$ \cite{huggins1941solutions,flory1942thermodynamics,tozzi2019out}.
The chemical potential includes the entropy of mixing, a maximum number density, $\Gamma_\infty$, and enzyme-enzyme interactions through $\chi$ and $\Lambda$. 
A negative $\chi$ corresponds to attractive enzyme-enzyme interactions, which can result in phase separation with coexisting regions of high and low surface concentration of enzymes \cite{baumgart2007large}. Enzymes and proteins suspended in solution often aggregate above a threshold concentration, which suggests some degree of attraction even when they lie on a surface. Here, we consider weak interactions so that at equilibrium, in the absence of chemical reactions, there is no phase separation and $\Gamma$ is homogeneous over the surface and equal to its average value $\Gamma = \Gamma_0$. The last term of the chemical potential accounts for nonlocal interactions between the enzymes. By penalizing lateral gradients of enzymes, it regularizes the boundaries between regions of high and low concentration of enzymes and it is mathematically required when $\chi<0$ \cite{illposed1984}.

We make Eqs.~(\ref{Laplacedim},\ref{enztransport}) dimensionless using $R$ as characteristic lengthscale, $R^2/D_s$ as characteristic time, $\Gamma_\infty$ as characteristic enzyme area density and $k_\text{cat} \Gamma_\infty R/D$ as characteristic product number density. By doing so, we find four dimensionless numbers. The P\'{e}clet number, $Pe = f b  k_\text{cat}R \Gamma_\infty /D \, D_s$, expresses the relative importance of advection and diffusion of enzymes over the surface. Since the phoretic mobility coefficient can be positive or negative, $Pe$ is also signed. The sign of $Pe$ indicates whether the enzymes are advected along or against the surface gradient of product. In experiments using enzymes \cite{ma2016motion} the characteristic phoretic velocity is $b k_\text{cat} \Gamma_\infty /D \approx 10 \, \rm{\upmu m \, s^{-1}}$, the surface diffusion of enzymes is usually slow $D_s \approx 1 \, \rm{\upmu m^2 s^{-1}}$, which result in $Pe$ of  $\mathcal{O}(1)$ for a $R \approx 1 \,\rm{\upmu m}$ and $f=0.1$. We note however that the P\'eclet number can be higher for larger particles.
The dimensionless enzyme-enzyme interaction is given by $\chi^* = \chi /k_BT \, \Gamma_\infty$. The dimensionless nonlocal enzyme-enzyme interaction parameter is defined as $\Lambda^* = \Lambda /k_BT \, R^2 \,\Gamma_\infty $ and it is always positive. 
Finally, the mean number density divided by the maximum density $\Gamma_0^*=\Gamma_0/\Gamma_\infty$ represents the degree of coverage of the surface and varies between zero and one.

It is straightforward to show that $\Gamma = \Gamma_0$ and $c = k_\text{cat} R^2 \Gamma_0/ D \, r$ are a solution to the Eqs. \eqref{Laplacedim}-\eqref{enztransport}, which corresponds to a spherically-symmetric distribution of $c$ around the particle. However, this solution can become unstable to infinitesimal fluctuations of the enzyme distribution.  
We study this phenomenon by performing a linear stability analysis of the homogeneous solution and fully nonlinear numerical simulations of Eqs.\eqref{Laplacedim}-\eqref{enztransport}. 

\begin{figure}[ht]
\centering
\includegraphics[width=0.5\textwidth]{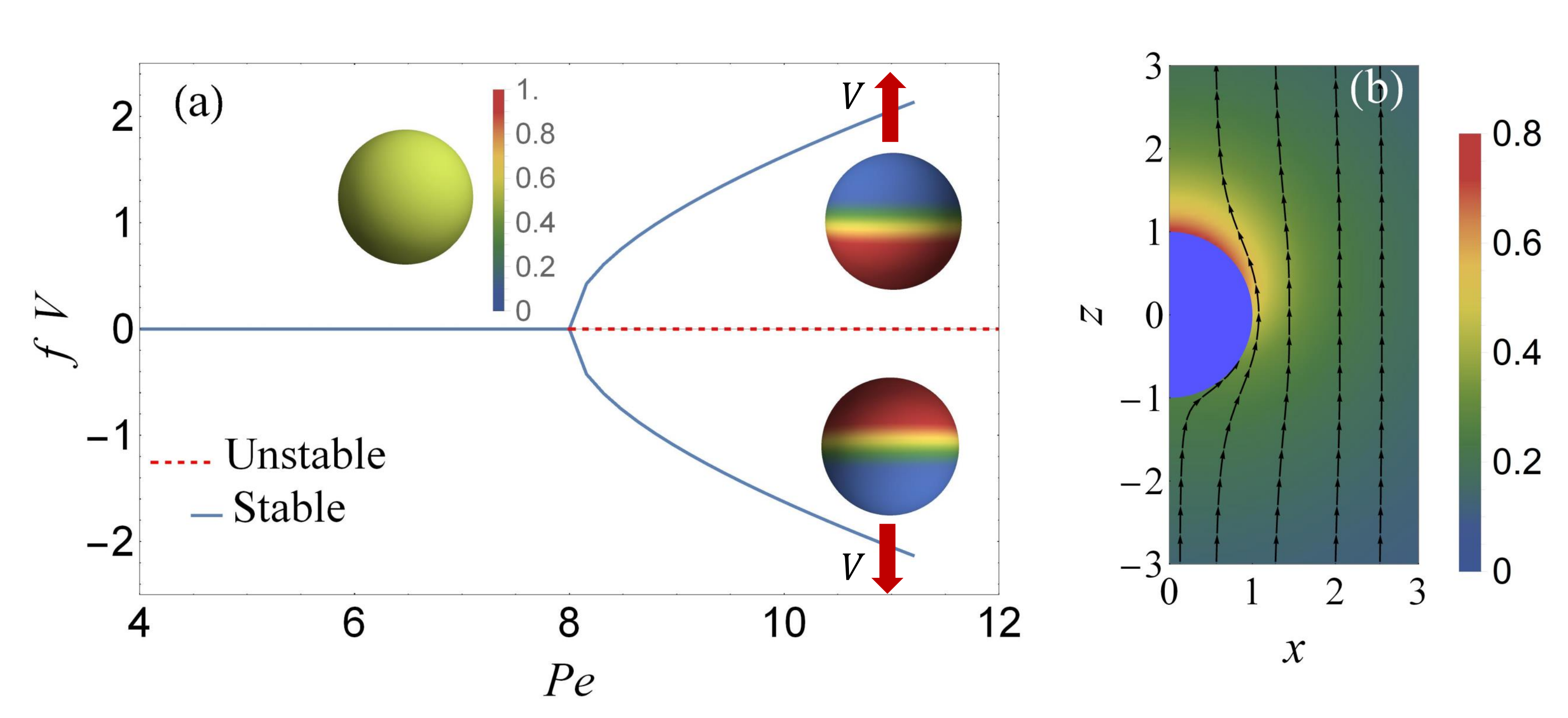}
\caption{Spontaneous polarization and locomotion: (a) The dimensionless velocity of the particle as a function of $Pe$ for $\Gamma_0^*=0.5$ and no enzyme-enzyme interaction $\chi^*=0$ and $\Lambda^*=0$. The insets show the dimensionless enzyme distribution in homogeneous and polarized states. (b) The dimensionless concentration of the product around the particle and the streamlines corresponding to $Pe = 11$. }
\label{fig3}
\end{figure}

We consider small axisymmetric perturbations $\Gamma = \Gamma_0 +\delta \Gamma $ and $c = k_\text{cat} R^2 \Gamma_0/ D r + \delta c$ about the homogeneous steady state and expand $\delta c$ and $\delta \Gamma$ in Legendre polynomials as $\delta c= \sum_{l=1}^{\infty}\, \delta c_l (0) \exp(\lambda_l \, t) \,  r^{-l-1} \, P_l \left(\cos{\theta}\right)$ $\delta \Gamma = \sum_{l=1}^{\infty} \, \delta \Gamma_l (0) \exp(\lambda_l \, t) \, P_l \left(\cos{\theta}\right)$, with $P_l \left(\cos{\theta}\right)$ the Legendre polynomial of degree $l$ and $\theta$ the polar angle. The expansions of $\delta c$ and $\delta \Gamma$ represent eigenfunctions of Eqs. \eqref{Laplacedim}-\eqref{enztransport}. $\delta \Gamma_l (0)$ and $\delta c_l (0)$ are the initial values of the Legendre modes of the perturbations and $\lambda_l$ are their growth rate. If the real part of $\lambda_l$ is positive, then any small perturbation of the mode $l$ grows exponentially and the system is linearly unstable. By plugging the expansions $\Gamma = \Gamma_0 +\delta \Gamma $ and $c = k_\text{cat} R^2 \Gamma_0/ D r + \delta c$ into the governing Eqs.~(\ref{Laplacedim},\ref{enztransport}) and keeping only the linear terms, we find that $\lambda_l$ is real and we obtain a relation between the dimensionless growth rate and the dimensionless parameters \cite{supmat}.

\begin{figure}[ht]
\centering
\includegraphics[width=0.45\textwidth]{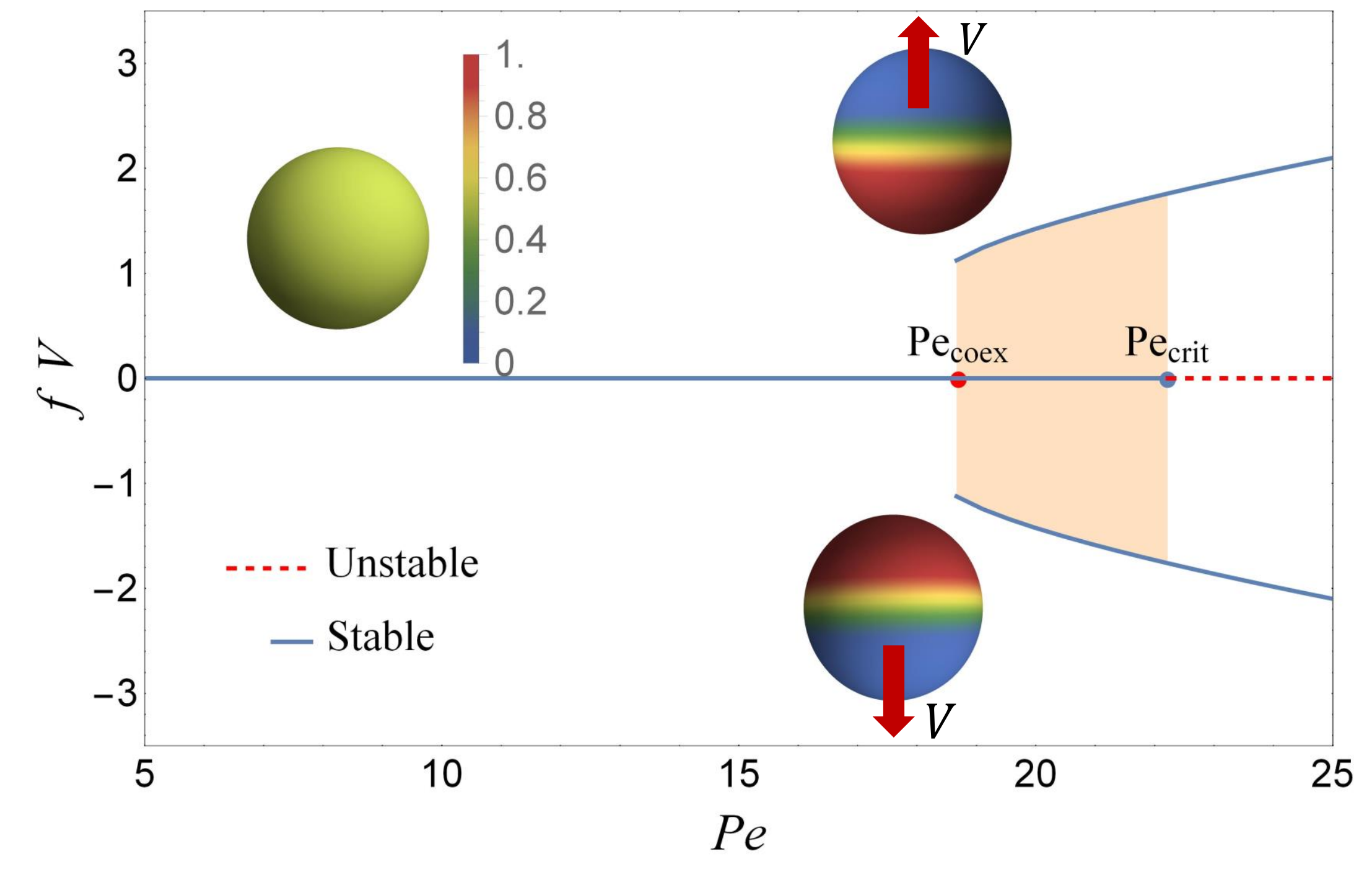}
\caption{The dimensionless velocity of the particle as a function of $Pe$ for $\Gamma_0^*=0.1$ and $\chi^*=0$ and $\Lambda^*=0$. The shaded area represents the region where stable homogeneous and polarized solutions coexist. $Pe_\text{coex}$ is obtained by numerical continuation of the stable polarized branch.} 
\label{fig4}
\end{figure}
The instability condition, $\lambda_l>0$, can be rewritten as a condition for the P\'{e}clet number, $Pe \, \Gamma_0^* > (l+1) \left[(1-\Gamma_0^*)^{-1}+\Gamma_0^* \, \chi^*+l(l+1) \Gamma_0^* \, \Lambda^*  \right] $ .
We restrict our study to cases where $\chi^* \geq -\left[(1-\Gamma_0^*)\Gamma_0^*\right]^{-1}$, for which the enzyme-enzyme interactions are weak and the equilibrium distribution of enzymes is homogeneous. In these cases, the right hand side of the inequality above is a monotonically increasing function of $l$ and the first unstable mode is the dipole, $l=1$. 
Thus, we define a critical P\'{e}clet number, 
\begin{equation}\label{pecritdipole}
  Pe_\text{crit}  =  \frac{2}{(1-\Gamma_0^*)\Gamma_0^*}+2\chi^*+4 \Lambda^* \, \, ,
\end{equation}
that discriminates homogeneous states that are stable from those that are unstable. Any homogeneous state with $Pe>Pe_\text{crit}$ is unstable to infinitesimal perturbations and spontaneously polarizes. Since $\chi^*>-\left[(1-\Gamma_0^*)\Gamma_0^*\right]^{-1}$ and $\Lambda^*>0$, the right hand side of Eq. \eqref{pecritdipole} is positive and only a positive $Pe$ leads to unstable steady states. In this case, phoretic flows advect more enzymes along the gradient of concentration and reinforce an initial perturbation as depicted in Fig. \ref{fig1}(b). Finally, Eq. \eqref{pecritdipole} shows that attractive enzyme-enzyme interactions, $\chi^*<0$, promote the instability by reducing $Pe_\text{crit}$.

To examine the distribution of enzymes and the particle velocity beyond the linear stability analysis, we resort to numerical simulations. We assume an axisymmetric solution and we expand the bulk and surface concentrations in Legendre modes, $c= \sum_{l=1}^{\infty}\,  c_l (t) \,  r^{-l-1} \, P_l \left(\cos{\theta}\right)$ and $\Gamma = \sum_{l=1}^{\infty} \, \Gamma_l (t) \, P_l \left(\cos{\theta}\right)$, and solve for the time-dependent coefficients $c_l (t)$ and $\Gamma_l (t)$ \cite{supmat}.  
In the supplementary material \cite{supmat} we show that the velocity of the particle is directly related to the dipolar mode, $\Gamma_1(t)$, as $\boldsymbol{V} = -\frac{Pe}{3 \, f} \Gamma_1(t) \hat{\boldsymbol{z}}$, with $\hat{\boldsymbol{z}}$ the unit vector along the $z$ axis. We perform numerical simulations starting from small perturbations around a homogeneous distribution of enzymes and compute the time evolution of the system until it reaches a steady state. 

In Fig. \ref{fig3}(a), we show the steady-state velocity of an active particle as a function of the P\'{e}clet number for $\Gamma_0^*=0.5$, no interactions between enzymes $\chi^*=0$, and $\Lambda^*=0$. As shown in the insets of Fig. \ref{fig3}a, at small P\'{e}clet numbers the homogeneous distribution of enzymes is stable and the particle does not move. The velocity undergoes a supercritical pitchfork bifurcation at $Pe=Pe_\text{crit}$ whereby the quiescent solution becomes unstable, the spherical symmetry breaks and two polarized steady states become stable. In fact, polarization can emerge in any direction but, without loss of generality, our parametrization of the solutions describes  only two of these directions. The critical P\'{e}clet number matches that predicted by the linear stability analysis. At $Pe>Pe_\text{crit}$, the spherical symmetry breaks, resulting in an asymmetry not only of the enzymes but also of the product, Fig. \ref{fig3}(b). 

The velocity streamlines are shown in Fig. \ref{fig3}(b) in the co-moving frame. Since the flow field is generated by a surface slip velocity, we can use the squirmer model \cite{lighthill1952squirming,blake1971spherical} to rationalize it. We find that, the particle does not exert a force dipole to the fluid and behaves as a neutral squirmer \cite{lauga2016stresslets}. Therefore, the far field velocity field decays quickly as $ r^{-3}$ in the far field. This finding has implications for the collective motion of multiple particles as hydrodynamic interactions might decay faster than particle-particle phoretic interactions \cite{agudo2019phase,liebchen2019interactions,saha2019pairing,singh2019competing,kanso2019phoretic,nasouri2020exact}.

As the average number density of enzymes is reduced from $\Gamma_0^*=0.5$ to $\Gamma_0^*=0.1$, the pitchfork bifurcation occurring at $Pe=Pe_\text{crit}$ changes from supercritical to subcritical. This is depicted in Fig.~\ref{fig4}, where we plot the velocity of the particle for $\Gamma_0^*=0.1$. In contrast to what we found for $\Gamma_0^*=0.5$, Fig.~\ref{fig4} shows that stable polarized and stable homogeneous solutions coexist for a range of P\'{e}clet numbers. As a consequence, a particle with an homogeneous distribution of enzymes suddenly jumps to a polarized state with a finite velocity, once $Pe>Pe_\text{crit}$. By increasing and then decreasing the P\'{e}clet number, the distribution of enzymes undergoes a hysteresis loop: a polarized state emerges for $Pe>Pe_\text{crit}$ and disappears for $Pe<Pe_\text{coex}$ with $Pe_\text{coex}<Pe_\text{crit}$. Such hysteresis loop might be observed in experiments where the P\'{e}clet number is tuned through the reaction rate or where an external field is used to drive the system from the homogeneous stable branch to the polarized one. 
\begin{figure*}[hbt]
\centering
\includegraphics[width=0.8\textwidth]{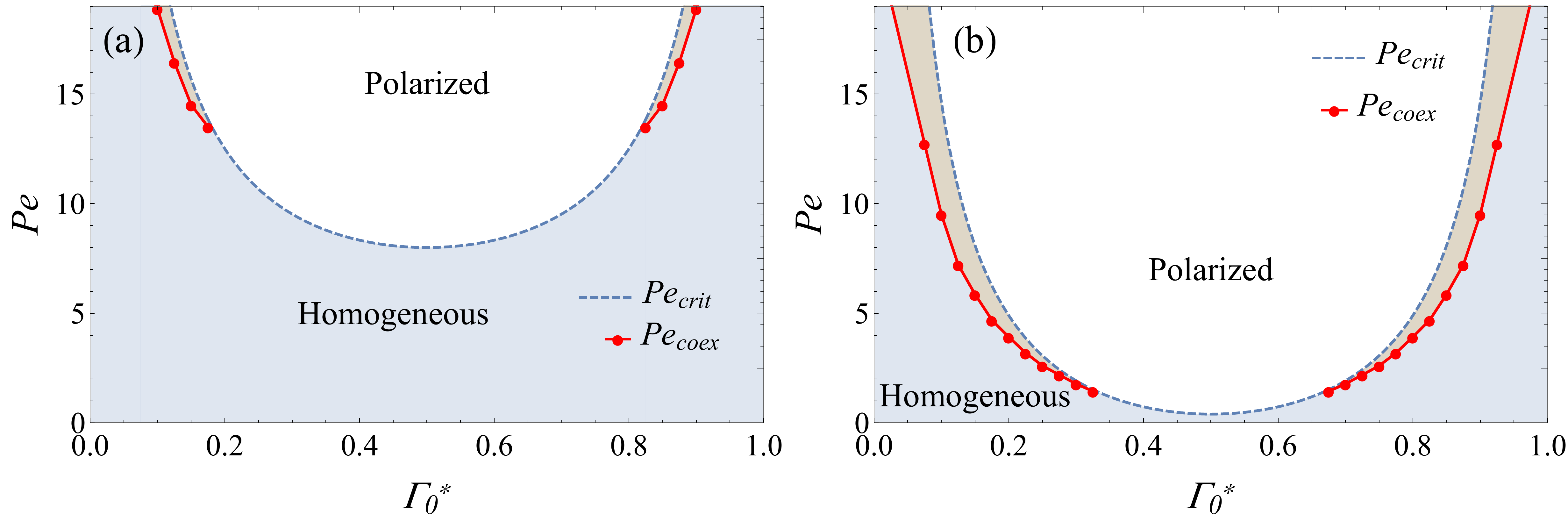}
\caption{Phase diagrams of the enzyme distribution in the case (a) no enzyme-enzyme interactions, $\chi^*=0$ and  $\Lambda^*=0$ and (b) attractive enzyme-enzyme interactions, $\chi^*=-4$ and  $\Lambda^*=0.1$. The blue-shaded area denotes regions where the homogeneous state is the only stable solution. In the white region the only stable solution is the polarized state. The orange-shaded region represents parameters for which the polarized and homogeneous solutions coexist.} 
\label{fig5}
\end{figure*}

The overall behavior of the active particle is summarized in Fig.~\ref{fig5}, where we map the regions where the homogeneous state, the polarized state, or both are stable depending on the average coverage of enzymes $\Gamma_0^*$ and $Pe$. We present such diagrams in the cases of no enzyme-enzyme interactions and of attractive enzyme-enzyme interactions. In both cases, polarized solutions can arise at lower $Pe$ for intermediate enzyme coverages. For non-interacting enzymes and the optimal coverage, the critical $Pe$ is about 8, Fig.~\ref{fig5}(a), which requires fast reaction kinetics, large phoretic mobility coefficient, large particle size and/or slow surface diffusion. In addition to these somewhat experimentally controllable knobs, an attractive self-interaction between enzymes of a few $k_BT$ can significantly expand the regions of coexistence and reduce $Pe_\text{crit}$, below one for optimal coverages, Figs.~\ref{fig5}(b).

Our results identify the key parameters that govern the active self-polarization of the particle. As shown by Eq. \eqref{pecritdipole} and by the definition of $Pe$, active self-polarization is favored by either decreasing the critical P\'{e}clet number, i.e. with intermediate enzyme coverage and slightly attractive enzyme-enzyme interactions, or by increasing $Pe$. The latter can be achieved in experiments by choosing bulky enzymes leading to a larger $f$, strong product-surface interactions leading to a large phoretic coefficient $b$, large catalytic rates $k_\text{cat}$ \cite{supmat}, large particles, and small enzyme mobility. If polarization takes place, then the velocity grows with $Pe$. 

There is some similarity between the spontaneous polarization under nonequilibrium conditions and the classic equilibrium liquid-liquid demixing \cite{jones2002soft}. For instance, Fig. \ref{fig5} shows that for small and large $\Gamma_0^*$ there is a region where a homogeneous solution coexists with a polarized one. In these regions, polarization occurs with a finite jump of the concentration dipole (and particle velocity), which is reminiscent of a first order phase transition. In the rest of the phase diagram, the transition to polarized states occurs with a continuous increase of the concentration dipole, which resembles a second order phase transition. However, since the mechanism that we discuss here is out-of-equilibrium requiring energy input from chemical reactions, the comparison with liquid-liquid demixing is only an analogy. 

In summary, we have identified a mechanism for the self-propulsion of chemically active particles, which rather than having hard-wired asymmetry, spontaneously develop active  polarization enabled by the lateral mobility of enzymes on their surface. In short, a perturbation of surface enzyme density in an otherwise uniformly coated particle results in an asymmetric reaction rate, which generates a gradient of product and phoretic flows along the particle surface. The advection of mobile enzymes over the surface by these  flows reinforces the initial perturbation, ultimately leading to a self-sustained polarization of the enzyme distribution and steady particle motion. This  mechanism bears similarity to other active mechano-chemical symmetry-breaking instabilities exploited by cells to divide, polarize or migrate \cite{RUPRECHT2015673,Bergert:2015aa,PhysRevLett.116.028102,mietke2019minimal}. Our results could be useful to design bio-mimicking active particles with an adaptive/controllable propulsion mechanism, which can be dynamically (dis-)engaged by sensing or tuning any of the physical parameters involved in the self-polarization instability, as mapped in the present study.

\begin{acknowledgments}
M.D.C. acknowledges funding from the European Union’s Horizon 2020 research and innovation program under the Marie Skłodowska-Curie action (GA 712754), the Severo Ochoa programme (SEV-2014-0425) and the CERCA Programme/Generalitat de Catalunya. I.P. acknowledges support from MINECO/FEDER Project No. PGC2018-098373-B-I00, DURSI Project No. 2017SGR-884, SNF Project No. 200021-175719 and the H2020 research and innovation program under the FET open project NanoPhlow. M. A. acknowledges the  support from the Generalitat de Catalunya (ICREA Academia Award, 2017-SGR-1278), from the Spanish Ministry of Economy and Competitiveness, through the Severo Ochoa Programme (CEX2018-000797-S), and from the European Research Council (CoG-681434). S.S. acknowledges the BBVA Foundation for the MEDIROBOTS project and the CERCA program by the Generalitat de Catalunya.
\end{acknowledgments}

\bibliography{Spntaneous_polarization_bib}

\end{document}